# EFFECT OF SIMULATED SPACE CONDITIONS ON FUNCTIONAL CONNECTIVITY

**Parshuram N Aarotale** and Jaydip Desai

Department of Biomedical Engineering, Wichita State University, Wichita, KS 67260, USA

**Corresponding Author:** Jaydip Desai, Wichita State University, 1845 Fairmount St, Wichita, KS 67260
Email: pnaarotale@shockers.wichita.edu, jaydip.desai@wichita.edu

## ABSTRACT

Long duration spaceflight missions can affect the cognitive and behavioral activities of astronauts due to changes in gravity. The microgravity significantly impacts the central nervous system physiology which causes the degradation in the performance and lead to potential risk in the space exploration. The aim of this study was to evaluate functional connectivity at simulated space conditions using an unloading harness system to mimic the body-weight distribution related to Earth, Mars, and International Space Station. A unity model with six directional arrows to imagine six different motor imagery tasks associated with arms and legs were designed for the Oculus Rift S virtual reality headset for testing. An Electroencephalogram (EEG) and functional near infrared spectroscopy (fNIRS) signals were recorded from 10 participants in the distributed weight conditions related to Earth, Mars, and International Space station using the g.Nautilus fNIRS system at sampling rate of 500 Hz. The magnitude squared coherence were estimated from left vs right hemisphere of the brain that represents functional connectivity. The EEG coherence was the higher which shows the strong functional connectivity and fNIRS coherence was lower shows weak functional connectivity between left vs right hemisphere of the brain, during all the tasks and trials irrespective of the simulated space conditions. Further analysis of functional connectivity needed between the intra-regions of the brain.

## INTRODUCTION

Over the last two decades, long-duration spaceflights were studied, showing that astronauts' cognitive and behavioral activities were impaired in space environments [1]. Long duration spaceflights cause changes in the sensorimotor systems which results in the difficulties experienced by astronauts with movements, postural control, and manual control. Microgravity is the leading cause of performance degradation in space exploration, and it poses a severe threat [2, 3]. Space flight and the microgravity environment significantly impacted astronaut central nervous system physiology [4]. This study evaluates human brain responses in distributed-weight conditions related to Earth, Mars, and International Space Station (ISS). This study used a harness system with a weight scale to distribute human body weight associated with Earth, Mars, and ISS. EEG and fNIRS signals were recorded to find the coherence during distributed weight conditions related to earth, mars, and ISS. EEG coherence measures the correlation coefficient that estimates the relative amplitude and phase consistency between any pair of signals at different frequency bands [5]. The functional connectivity was calculated by finding out the EEG and fNIRS coherence from the brain's left vs. right hemisphere.

## MATERIALS & METHODS

Guger Technologies' g.Nautilus fNIRS wireless system was utilized because of its wireless capability and dual modality (EEG and fNIRS). The 500 Hz sampling rate was utilized during recording the signals. Figure 1 shows an unloading harness system where a participant is standing straight on the floor, partially lifted, and completely lifted to mimic body-weight distribution related to Earth, Mars, and ISS respectively.

Data collection and study protocol

Virtual reality technology was integrated in this study for Motor Imaginary (MI) tasks. An Oculus Rift S VR headset was used to give three-dimensional view of the MI stimulation. Unity model was designed to generate the MI tasks which contain six arrows that appears on the VR display. The designed Unity model randomly picks the arrow and flashes it for 3 seconds. There is a rest period (3 seconds) between simulations.

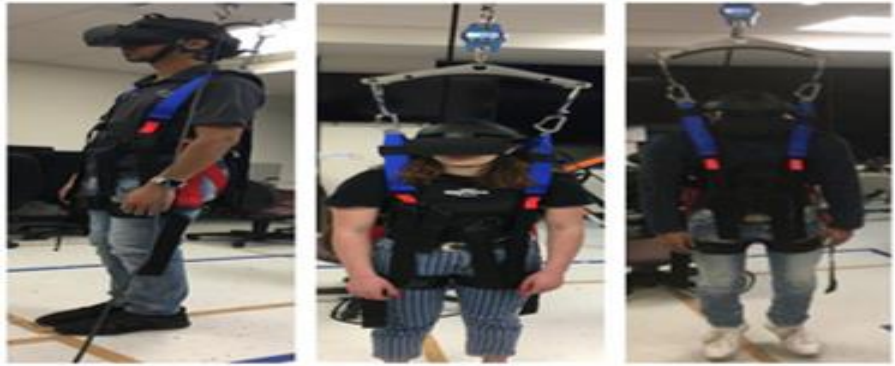

*Figure 1: Distributed-Weight Environment a) Standing –Earth b) Distributed weight-Mars, and c) Completely Lifted-ISS.*

A total of 10 adults were recruited (5 male and 5 females all aged between 21 and 30 years). The experiment and the procedure were approved by the Institutional Review Board (IRB). None of the selected participants had participated in any previous brain signal acquisition experiment, and none had a history of any psychiatric, neurological, or visual disorder. All of them had normal or corrected to normal vision, and all provided a written consent after having been informed in detail about the experimental procedure. Testing on participants was temporarily halted due to the COVID-19 restrictions and was resumed only after a clear go-ahead was issued by the IRB with the following measures in place. Various measures like mandatory mask policy, contact tracing of the participants and the lab members until two weeks after the testing period of the experiment and thorough cleaning of all the equipment handled during the test was done to protect the health of participants as well as the researchers in the lab. Figure 2 shows the montage system used for brain signal acquisition and a neuro-virtual interface system.

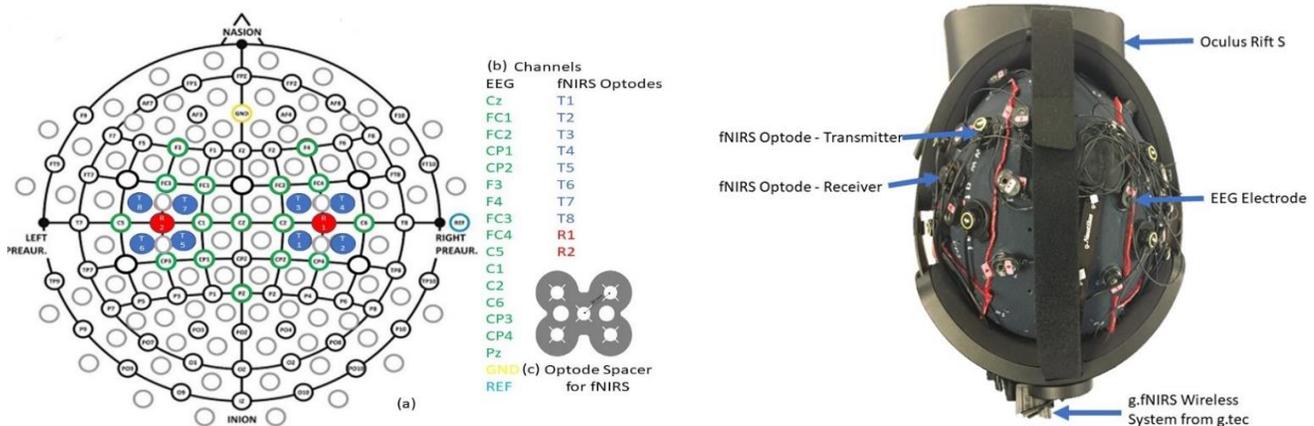

*Figure 2: Montage system for EEG and fNIRS system a) Montage system-based on International 10-20 System, b) Sixteen EEG and 8 fNIRS Channels to cover motor and somatosensory cortex regions, and c) fNIRS optode spacer with 30mm distance from transmitter to receiver d) Neuro-Virtual Interface.*

EEG and fNIRS Data Processing

Raw EEG signals were (preprocessed) low pass filtered with cut off frequency of 30 Hz using finite impulse response (FIR) filter. This filter is designed using the window method with 50 percentage overlap hamming windows. The magnitude squared coherence from Left vs Right hemisphere EEG channels were extracted in the frequency bands [13-30] HZ. FNIRS analysis were based on total hemoglobin concentrations (HBT=HBO+HBR) amplitude. The Left vs Right hemisphere fNIRS coherence were extracted to find functional connectivity among these regions.

Coherence

This study focuses on changes in coherence in the of inter-region hemispheres, magnitude squared coherence Cxy of signals x and y was estimated by using Power spectral density (Pxx, Pyy) and cross Power spectral density (Pxy), It is expressed as

$$Cxy(f) = \frac{|Pxy(f)|^2}{Pxx(f) * Pyy(f)}$$

(eq. 1)

Where f is the frequency. The Welch's averaged method was utilized to calculate the coherence[6][7]. The coherence for frequency band [13-30Hz] was obtained from EEG.

## RESULTS

fNIRS Coherence

The magnitude squared coherence (MS Coherence) obtained from (Trial3, Figure 3) shows the changes in coherence between left hemisphere vs right hemisphere of the brain. The results were shown for different motor imaginary tasks (left, right, up, down, in, and out arrows). During the up-imaginary task (both hands), coherence values between the left and right hemispheres ranged from 0.35-0.57 for MIHS, 0.33-0.59 for MIHM, and 0.34-0.55 for MIHNL. During down-imaginary task (both feet), coherence values between the left and right hemispheres ranged from 0.33-0.57 for MIHS, 0.33-0.57 for MIHM, and 0.37-0.52 for MIHNL. During left-imaginary task (left hand), coherence values ranged from 0.29-0.57 for MIHS, 0.34-0.57 for MIHM, and 0.36-0.59 for MIHNL. During right-imaginary task (right hand), coherence values ranged from 0.36-0.56 for MIHS, 0.32-0.54 for MIHM, and 0.38-0.58 for MIHNL. During in-imagery task (left hand and left foot), coherence values ranged from 0.35 to 0.54 for MIHS, 0.33 to 0.60 for MIHM, and 0.39 to 0.54 for MIHNL. During out-imaginary task (right hand and right foot), coherence values ranged from 0.32-0.58 for MIHS, 0.28-0.48 for MIHM, and 0.36-0.55 for MIHNL. Lower coherence values indicates that no correlation between left and right hemisphere of brain. Comparing coherence between MIHM vs MIHS and MIHNL vs MIHS during all six different tasks shows no significant variation among all brain regions of left vs right hemisphere as compared to MIHS which determines that human brain has ability to adapt to distributed weight conditions. However, it is important to note that this study was limited to only five stimulations for each motor imagery tasks. The mean values fNIRS coherence for all the trials were shown in the table1. It is important to note that t2-t6 and t4-t8 channels were showing higher coherence compared with t1-t5 and t3-t7 channels regardless of any imaginary tasks.

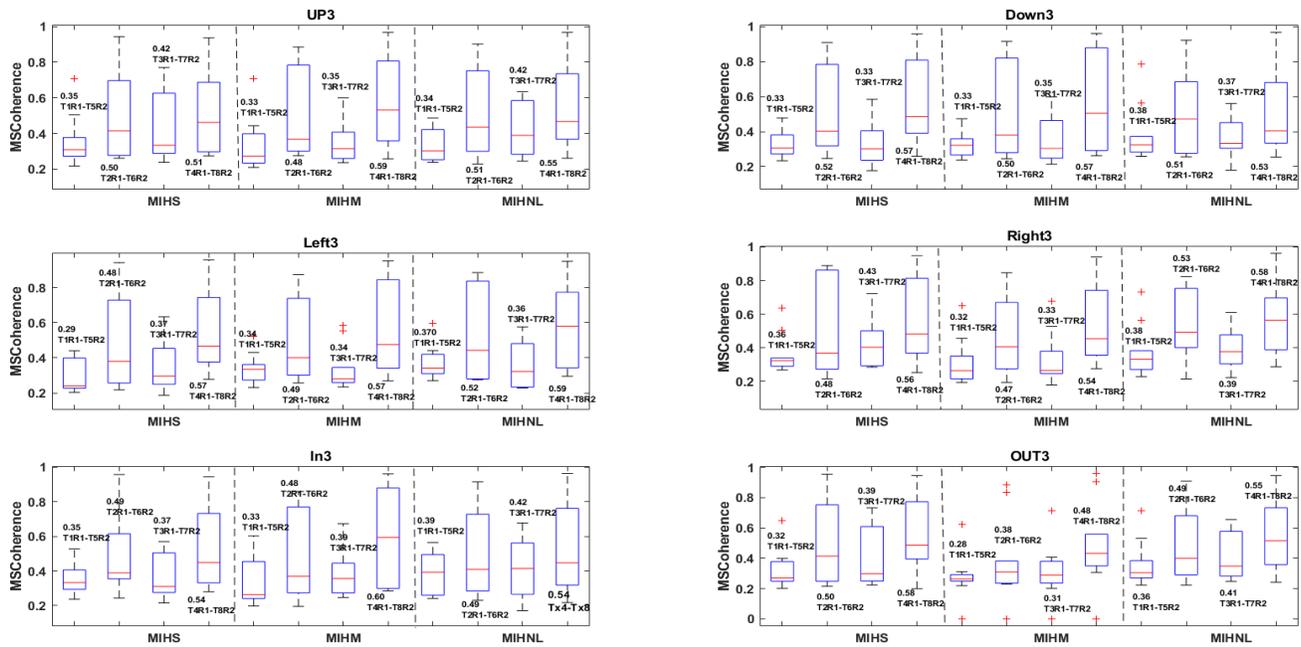

*Figure3: fNIRS Magnitude squared coherence for Trial3. The coherence values are lower at all different Motor imaginary tasks irrespective of distributed weight conditions at earth, Mars and ISS.*

*Table 1: Mean values of the fNIRS Coherence for all trials.*

|  | **MIHS** | | | | **MIHM** | | | | **MIHNL** | | | |
|---|---|---|---|---|---|---|---|---|---|---|---|---|
|  | t1-t5 | t2-t6 | t3-t7 | t4-t8 | t1-t5 | t2-t6 | t3-t7 | t4-t8 | t1-t5 | t2-t6 | t3-t7 | t4-t8 |
| UP1 | 0.32 | 0.49 | 0.40 | 0.56 | 0.33 | 0.49 | 0.34 | 0.60 | 0.38 | 0.48 | 0.40 | 0.59 |
| Down1 | 0.34 | 0.56 | 0.42 | 0.55 | 0.30 | 0.47 | 0.32 | 0.53 | 0.35 | 0.52 | 0.38 | 0.56 |
| Left1 | 0.37 | 0.49 | 0.40 | 0.60 | 0.31 | 0.49 | 0.36 | 0.56 | 0.33 | 0.48 | 0.39 | 0.55 |
| Right1 | 0.39 | 0.54 | 0.36 | 0.55 | 0.33 | 0.48 | 0.36 | 0.55 | 0.34 | 0.50 | 0.38 | 0.56 |
| IN1 | 0.33 | 0.49 | 0.38 | 0.55 | 0.37 | 0.47 | 0.34 | 0.55 | 0.32 | 0.49 | 0.30 | 0.54 |
| Out1 | 0.33 | 0.52 | 0.39 | 0.56 | 0.32 | 0.41 | 0.34 | 0.47 | 0.37 | 0.45 | 0.41 | 0.56 |
| UP3 | 0.35 | 0.50 | 0.42 | 0.51 | 0.33 | 0.48 | 0.35 | 0.59 | 0.34 | 0.51 | 0.42 | 0.55 |
| Down3 | 0.33 | 0.52 | 0.33 | 0.57 | 0.33 | 0.50 | 0.35 | 0.57 | 0.38 | 0.51 | 0.37 | 0.53 |
| Left3 | 0.29 | 0.48 | 0.37 | 0.57 | 0.34 | 0.49 | 0.34 | 0.57 | 0.37 | 0.52 | 0.36 | 0.59 |
| Right3 | 0.36 | 0.48 | 0.43 | 0.56 | 0.32 | 0.47 | 0.33 | 0.54 | 0.38 | 0.53 | 0.39 | 0.58 |
| IN3 | 0.35 | 0.49 | 0.37 | 0.54 | 0.33 | 0.48 | 0.39 | 0.60 | 0.39 | 0.49 | 0.42 | 0.54 |
| Out3 | 0.32 | 0.50 | 0.39 | 0.58 | 0.28 | 0.38 | 0.31 | 0.48 | 0.36 | 0.49 | 0.41 | 0.55 |
| UP5 | 0.34 | 0.49 | 0.37 | 0.53 | 0.30 | 0.49 | 0.36 | 0.60 | 0.39 | 0.52 | 0.39 | 0.53 |
| Down5 | 0.35 | 0.56 | 0.45 | 0.61 | 0.32 | 0.52 | 0.32 | 0.63 | 0.39 | 0.54 | 0.34 | 0.59 |
| Left5 | 0.37 | 0.50 | 0.42 | 0.55 | 0.29 | 0.44 | 0.31 | 0.54 | 0.42 | 0.52 | 0.37 | 0.59 |
| Right5 | 0.32 | 0.46 | 0.39 | 0.54 | 0.34 | 0.44 | 0.35 | 0.54 | 0.35 | 0.50 | 0.40 | 0.55 |
| IN5 | 0.35 | 0.56 | 0.39 | 0.61 | 0.30 | 0.48 | 0.35 | 0.54 | 0.39 | 0.52 | 0.37 | 0.56 |
| Out5 | 0.35 | 0.47 | 0.38 | 0.50 | 0.29 | 0.34 | 0.25 | 0.39 | 0.34 | 0.49 | 0.36 | 0.51 |

EEG Coherence

The Magnitude Squared coherence (MS Coherence) (Trial3, Figure4) shows the changes in coherence between inter regions (left hemisphere vs right hemisphere) of brain. The results were shown for different motor imaginary tasks (left, right, up, down, in, and out). For up task, during MIHS the coherence for frontal Central (0.99, 0.89), central parietal (0.96, 0.91), central region (0.95, 0.83) and frontal region (0.88) was higher; indicates high correlation between left and right hemisphere of brain. During MIHM the coherence for fontal central (0.97, 0.90), central parietal (0.96, 0.92), central (0.95, 0.81) and frontal region (0.92) were higher indicating high correlation between left and right hemisphere of brain. During MIHNL, the coherence for fontal Central (0.99, 0.90), central parietal (0.95, 0.90), central (0.94, 0.76) and frontal region (0.91) were higher indicating high correlation between left and right hemisphere of brain. For Down task, during MIHS the coherence for frontal Central (0.99, 0. 9), central parietal (0.96, 0.9), central region (0.95, 0.81) and frontal region (0.90) was higher; indicates high correlation between left and right hemisphere of brain. During MIHM the coherence for fontal Central (0.97, 0.90), central parietal (0.90, 0.96), central (0.93, 0.95) and frontal region (0.82) were higher indicating high correlation between left and right hemisphere of brain. During MIHNL, the coherence for fontal Central (0.99, 0.90), central parietal (0.95, 0.90), central (0.95, 0.79) and frontal region (0.93) were higher indicating high correlation between left and right hemisphere of brain.

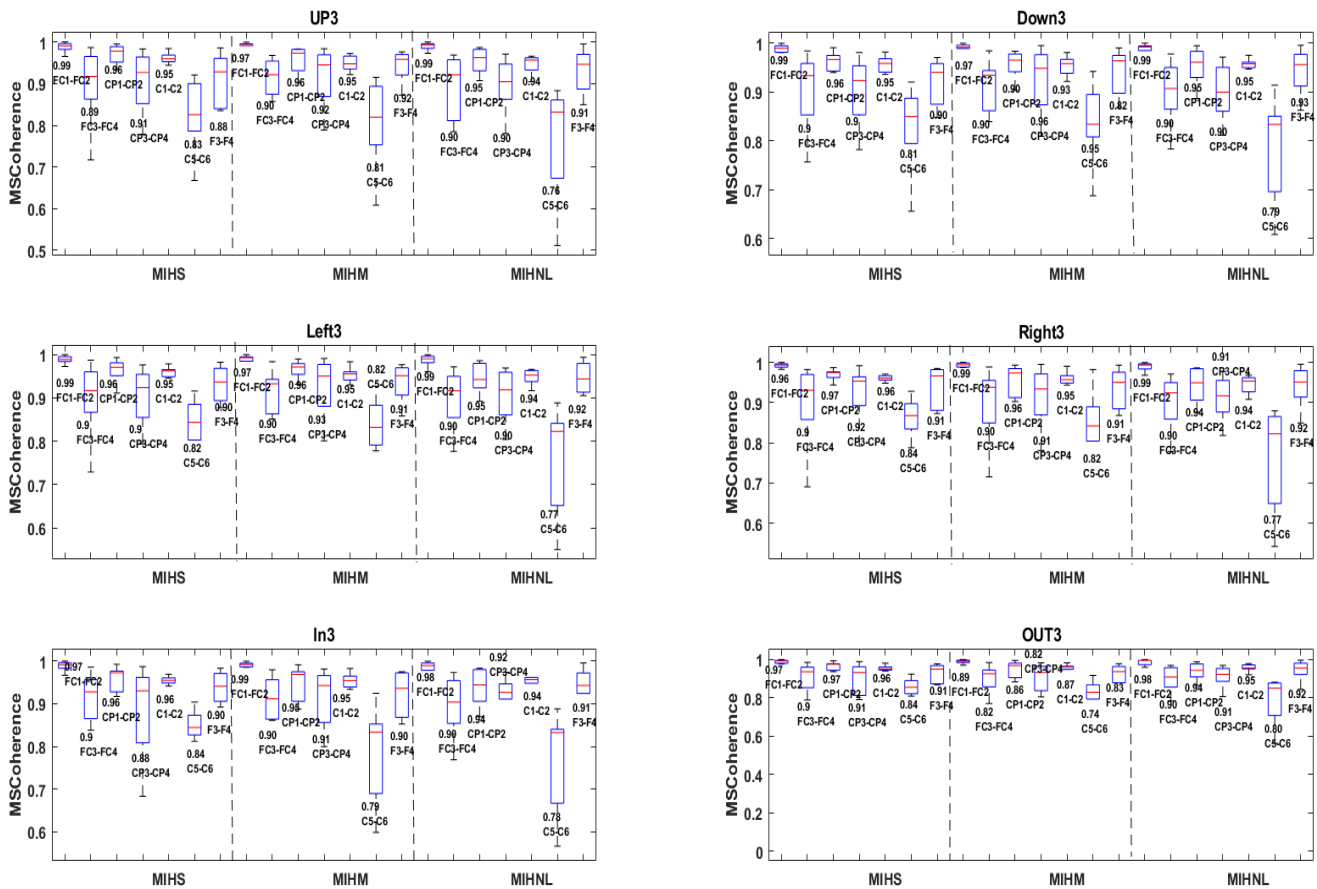

Figure 4: EEG Magnitude squared coherence for Trial3. The coherence values are higher at all different Motor imaginary tasks irrespective of distributed weight conditions at earth, Mars and ISS.

For left task, during MIHS the coherence for frontal Central (0.99, 0. 9), central parietal (0.96, 0.9), central region (0.95, 0.82) and frontal region (0.90) was higher; indicates high correlation between left and right hemisphere of brain. During MIHM the coherence for fontal Central (0.97, 0.90), central parietal (0.96, 0.93), central (0.95, 0.82) and frontal region (0.91) were higher indicating high correlation between left and right hemisphere of brain. During MIHNL, the coherence for fontal Central (0.99, 0.90), central parietal (0.95, 0.90), central (0.94, 0.77) and frontal region (0.92) were higher indicating high correlation between left and right hemisphere of brain. During right task, for MIHS the coherence for frontal Central (0.96, 0.9), central parietal (0.97, 0.92), central region (0.96, 0.84) and frontal region (0.91) was higher; indicates high correlation between left and right hemisphere of brain. During MIHM the coherence for fontal Central (0.99, 0.90), central parietal (0.96, 0.91), central (0.95, 0.82) and frontal region (0.91) were higher indicating high correlation between left and right hemisphere of brain. During MIHNL, the coherence for fontal Central (0.99, 0.90), central parietal (0.94, 0.91), central (0.94, 0.77) and frontal region (0.92) were higher indicating high correlation between left and right hemisphere of brain.

For in task, during MIHS the coherence for frontal Central (0.97, 0.9), central parietal (0.96, 0.88), central region (0.96, 0.84) and frontal region (0.90) was higher; indicates high correlation between left and right hemisphere of brain. During MIHM the coherence for fontal Central (0.99, 0.90), central parietal (0.95, 0.91), central (0.95, 0.79) and frontal region (0.90) were higher indicating high correlation between left and right hemisphere of brain. During MIHNL, the coherence for fontal Central (0.98, 0.99), central parietal (0.94, 0.92), central (0.94, 0.78) and frontal region (0.91) were higher indicating high correlation between left and right hemisphere of brain. For out task, during MIHS the coherence for frontal Central (0.97, 0.9), central parietal (0.97, 0.91), central region (0.96, 0.84) and frontal region (0.91) was higher; indicates high correlation between left and right hemisphere of brain. During MIHM the coherence for fontal Central (0.89, 0.82), central parietal (0.86, 0.82), central (0.87, 0.74) and frontal region (0.83) were higher indicating high correlation between left and right hemisphere of brain. During MIHNL, the coherence for fontal Central (0.98, 0.90), central parietal (0.94, 0.91), central (0.95, 0.80) and frontal region (0.92) were higher indicating high correlation between left and right hemisphere of brain. Comparing coherence between MIHM vs MIHS and MIHNL vs MIHS shows no significant variation among brain regions of left and right hemisphere as compared to MIHS for all the six different tasks. The mean values EEG coherence for the trials 1 and 5 were shown in the table2.

Table 2: Mean values of the EEG Coherence for recording trial1 and 5.

|      | Regions | Up1  | Down1 | Left1 | Right1 | In1  | Out1 | Up5  | Down5 | Left5 | Right5 | In5  | Out5 |
|------|---------|------|-------|-------|--------|------|------|------|-------|-------|--------|------|------|
|      | Fc1-Fc2 | 0.99 | 0.99  | 0.99  | 0.99   | 0.99 | 0.99 | 0.99 | 0.98  | 0.98  | 0.97   | 0.97 | 0.99 |
|      | Fc3-Fc4 | 0.89 | 0.9   | 0.91  | 0.92   | 0.91 | 0.91 | 0.89 | 0.9   | 0.9   | 0.9    | 0.9  | 0.9  |
|      | Cp1-Cp2 | 0.92 | 0.96  | 0.96  | 0.97   | 0.96 | 0.95 | 0.96 | 0.95  | 0.96  | 0.96   | 0.96 | 0.96 |
|      | Cp3-Cp4 | 0.9  | 0.92  | 0.9   | 0.93   | 0.92 | 0.91 | 0.91 | 0.9   | 0.91  | 0.91   | 0.91 | 0.9  |
|      | C1-C2   | 0.94 | 0.95  | 0.94  | 0.96   | 0.95 | 0.95 | 0.95 | 0.96  | 0.95  | 0.96   | 0.96 | 0.95 |
|      | C5-C6   | 0.82 | 0.83  | 0.79  | 0.85   | 0.84 | 0.82 | 0.81 | 0.84  | 0.82  | 0.84   | 0.83 | 0.81 |
| MIHS | F3-F4   | 0.9  | 0.89  | 0.91  | 0.91   | 0.89 | 0.9  | 0.9  | 0.88  | 0.9   | 0.91   | 0.92 | 0.89 |
|      | Fc1-Fc2 | 0.97 | 0.97  | 0.98  | 0.98   | 0.98 | 0.89 | 0.98 | 0.98  | 0.99  | 0.97   | 0.99 | 0.89 |
|      | Fc3-Fc4 | 0.91 | 0.9   | 0.91  | 0.91   | 0.9  | 0.81 | 0.91 | 0.92  | 0.89  | 0.9    | 0.9  | 0.81 |
|      | Cp1-Cp2 | 0.96 | 0.93  | 0.94  | 0.94   | 0.94 | 0.85 | 0.96 | 0.97  | 0.95  | 0.94   | 0.95 | 0.87 |
|      | Cp3-Cp4 | 0.93 | 0.93  | 0.92  | 0.92   | 0.93 | 0.83 | 0.93 | 0.93  | 0.91  | 0.93   | 0.91 | 0.83 |
| MIHM | C1-C2   | 0.95 | 0.95  | 0.94  | 0.95   | 0.95 | 0.86 | 0.96 | 0.96  | 0.94  | 0.96   | 0.94 | 0.86 |

|  | C5-C6 | 0.82 | 0.8 | 0.79 | 0.8 | 0.79 | 0.75 | 0.84 | 0.84 | 0.78 | 0.82 | 0.78 | 0.72 |
|---|---|---|---|---|---|---|---|---|---|---|---|---|---|
|  | F3-F4 | 0.91 | 0.89 | 0.9 | 0.9 | 0.89 | 0.81 | 0.91 | 0.93 | 0.91 | 0.91 | 0.9 | 0.83 |
|  | Fc1-Fc2 | 0.99 | 0.99 | 0.99 | 0.98 | 0.99 | 0.99 | 0.99 | 0.88 | 0.99 | 0.99 | 0.98 | 0.99 |
|  | Fc3-Fc4 | 0.91 | 0.91 | 0.92 | 0.91 | 0.92 | 0.91 | 0.9 | 0.82 | 0.89 | 0.9 | 0.9 | 0.9 |
|  | Cp1-Cp2 | 0.95 | 0.95 | 0.95 | 0.95 | 0.95 | 0.95 | 0.95 | 0.86 | 0.94 | 0.94 | 0.95 | 0.94 |
|  | Cp3-Cp4 | 0.91 | 0.91 | 0.92 | 0.91 | 0.91 | 0.92 | 0.91 | 0.81 | 0.91 | 0.91 | 0.9 | 0.91 |
|  | C1-C2 | 0.94 | 0.95 | 0.94 | 0.95 | 0.94 | 0.95 | 0.94 | 0.85 | 0.95 | 0.94 | 0.95 | 0.94 |
|  | C5-C6 | 0.77 | 0.78 | 0.79 | 0.8 | 0.79 | 0.81 | 0.77 | 0.7 | 0.78 | 0.77 | 0.78 | 0.77 |
| MIHNL | F3-F4 | 0.92 | 0.92 | 0.93 | 0.91 | 0.91 | 0.92 | 0.92 | 0.86 | 0.92 | 0.91 | 0.92 | 0.92 |

## DISCUSSION

To measure the similarities between the electrical cortical activity of the electrodes, EEG and fNIRS coherence was estimated. The presence of the EEG coherence is the sign of interaction and synchronization among the brain regions. High coherence represents the measure of strong connectivity and low coherence represents the measure of weak connectivity [8].The obtained results for EEG coherence estimated between the left Vs right hemisphere of the brain was higher for each of MIHS,MIHM and MIHNL during all the trials of recording (figure 4,table2),reflecting the strong structural and functional connectivity among the regions of the brain. The functional connectivity was higher irrespective of simulated distributed weight condition related to earth, mars, and ISS. This reflects no major changes in the functional connectivity comparing MIHM Vs MIHS and MIHNL vs MIHM with respect to MIHS as baseline. The results for fNIRS coherence estimated between the left Vs right hemisphere of the brain was lower for each of MIHS, MIHM and MIHNL during all the trials of recording (figure 3, table1), reflecting the weak structural and functional connectivity among the regions of the brain. The functional connectivity was weak irrespective of simulated distributed weight condition related to earth, mars, and ISS. This reflects no major changes in the functional connectivity comparing MIHM Vs MIHS and MIHNL vs MIHM with respect to MIHS as baseline.

The signal quality of the recorded data can be improved by increasing the number of trials of recording on each participant and ensuring the proper placement of electrodes on participants. For the fNIRS data, the thickness of hair of the participants affects the hemodynamic response recorded hence affecting the integrity of the data. The electrodes that record the EEG activity are also affected by hair thickness, as these electrodes need to be placed close to the scalp and not doing so can either weaken the signal or produce signals with a lot of noise. The hair thickness also makes the optodes' cap fit incorrectly, causing issues such as increased space between the cap and the scalp depending on how thick the participant's hair is. For this participant, extra hairs were attempted to be moved aside around the cap once the optodes were in place. The gel for the EEG electrodes needs to be in direct contact with the scalp to get accurate readings. Other factors that affect the EEG recording include continuous eye movements, which may produce signal artifacts. Producing better quality signals can be achieved by informing the participant not to move and to stay relaxed, as any movements can affect the quality of the signal and measuring impedance of the electrodes both before and during the data recording. Further investigation needed for measuring functional connectivity among the intra-brain regions during distributed weight environment related earth, Mars, and ISS.

# CONCLUSIONS

The aim of the study was to evaluate functional connectivity by estimating the coherence between the inter brain regions during simulated space conditions. The EEG and fNIRS data were recorded in the distributed weight environment related to earth, Mars, and ISS. The harness system was utilized to mimic the simulated space conditions related to earth, Mars, and ISS. The virtual Reality (VR) headset was used, and unity model was created to display the six different tasks (Left, Right, Up, Down, IN and Out). The EEG coherence results shows the strong functional connectivity and fNIRS coherence shows the weak functional connectivity among the left vs right hemisphere of the brain during the simulated space conditions related earth, Mars, and ISS. The results shows that human brain can easily adapt to short duration weight distributed environment conditions. Advance research is necessary to understand human brain performance during long-duration space missions specifically mission critical tasks where astronauts are dealing with gravity transitions or changes into gravity while performing mission critical tasks. The further analysis of functional connectivity between the intra brain regions can be useful in selecting optodes' placement to obtain cognitive performance of human brain.

# ACKNOWLEDGMENTS

This research is funded by National Aeronautics and Space Administration (NASA) EPSCoR Research Infrastructure Development Grant 80NSSC19M0042. Authors would like to thank all the participants for their time and support.

# DISCLOSURES

All authors have nothing to declare.
**Institutional Review Board Approval #4460**
Wichita State University
1845 Fairmount St, Wichita, KS 67260, USA